\documentclass[11pt,a4paper]{article}
\usepackage{jcappub}

\usepackage{rotating}
\usepackage{booktabs} 
\usepackage{multirow} 
\usepackage{lineno}
\usepackage{xcolor}
\usepackage{orcidlink}
\usepackage[normalem]{ulem}
\usepackage{makecell}
\usepackage[draft]{changes} 
\definechangesauthor[name={RC}, color=red]{RC}
\usepackage{comment}
\usepackage{hyperref}

\title{Inference of the linear matter power spectrum at $z=0$ using DESI DR1 Full-Shape data}

\collaboration{DESI collaboration}

\author[1]{{R.~Cereskaite},}
\author[1]{{E.~Mueller},}
\author[2]{{C.~Howlett}\orcidlink{0000-0002-1081-9410},}
\author[2]{{Tamara~M.~Davis}\orcidlink{0000-0002-4213-8783},}
\author[3]{{J.~Aguilar},}
\author[4]{{S.~Ahlen}\orcidlink{0000-0001-6098-7247},}
\author[5,6]{{D.~Bianchi}\orcidlink{0000-0001-9712-0006},}
\author[7]{{D.~Brooks},}
\author[8,9]{{F.~J.~Castander}\orcidlink{0000-0001-7316-4573},}
\author[3]{{T.~Claybaugh},}
\author[3]{{A.~Cuceu}\orcidlink{0000-0002-2169-0595},}
\author[10]{{A.~de la Macorra}\orcidlink{0000-0002-1769-1640},}
\author[3,11]{{S.~Ferraro}\orcidlink{0000-0003-4992-7854},}
\author[12]{{A.~Font-Ribera}\orcidlink{0000-0002-3033-7312},}
\author[13,14]{{J.~E.~Forero-Romero}\orcidlink{0000-0002-2890-3725},}
\author[8,15,9]{{E.~Gaztañaga}\orcidlink{0000-0001-9632-0815},}
\author[16]{{G.~Gutierrez},}
\author[17]{{C.~Hahn}\orcidlink{0000-0003-1197-0902},}
\author[18,19,20]{{K.~Honscheid}\orcidlink{0000-0002-6550-2023},}
\author[21,22]{{D.~Huterer}\orcidlink{0000-0001-6558-0112},}
\author[23]{{M.~Ishak}\orcidlink{0000-0002-6024-466X},}
\author[24]{{R.~Joyce}\orcidlink{0000-0003-0201-5241},}
\author[24]{{S.~Juneau}\orcidlink{0000-0002-0000-2394},}
\author[25]{{D.~Kirkby}\orcidlink{0000-0002-8828-5463},}
\author[3]{{A.~Kremin}\orcidlink{0000-0001-6356-7424},}
\author[7]{{O.~Lahav},}
\author[3]{{A.~Lambert},}
\author[3]{{M.~Landriau}\orcidlink{0000-0003-1838-8528},}
\author[26]{{L.~Le~Guillou}\orcidlink{0000-0001-7178-8868},}
\author[27,12]{{M.~Manera}\orcidlink{0000-0003-4962-8934},}
\author[24]{{A.~Meisner}\orcidlink{0000-0002-1125-7384},}
\author[28,12]{{R.~Miquel},}
\author[29]{{J.~Moustakas}\orcidlink{0000-0002-2733-4559},}
\author[15]{{S.~Nadathur}\orcidlink{0000-0001-9070-3102},}
\author[30]{{J.~ A.~Newman}\orcidlink{0000-0001-8684-2222},}
\author[31,3]{{N.~Palanque-Delabrouille}\orcidlink{0000-0003-3188-784X},}
\author[32,33,34]{{W.~J.~Percival}\orcidlink{0000-0002-0644-5727},}
\author[35]{{F.~Prada}\orcidlink{0000-0001-7145-8674},}
\author[36]{{I.~P\'erez-R\`afols}\orcidlink{0000-0001-6979-0125},}
\author[37]{{G.~Rossi},}
\author[38]{{E.~Sanchez}\orcidlink{0000-0002-9646-8198},}
\author[3]{{D.~Schlegel},}
\author[21,22]{{M.~Schubnell},}
\author[39]{{H.~Seo}\orcidlink{0000-0002-6588-3508},}
\author[3]{{J.~Silber}\orcidlink{0000-0002-3461-0320},}
\author[24]{{D.~Sprayberry},}
\author[22]{{G.~Tarl\'{e}}\orcidlink{0000-0003-1704-0781},}
\author[24]{{B.~A.~Weaver},}
\author[26]{{P.~Zarrouk}\orcidlink{0000-0002-7305-9578},}
\author[3]{{R.~Zhou}\orcidlink{0000-0001-5381-4372},}
\author[40]{{H.~Zou}\orcidlink{0000-0002-6684-3997}}

\affiliation[1]{Department of Physics and Astronomy, University of Sussex, Brighton BN1 9QH, U.K}
\affiliation[2]{School of Mathematics and Physics, University of Queensland, Brisbane, QLD 4072, Australia}
\affiliation[3]{Lawrence Berkeley National Laboratory, 1 Cyclotron Road, Berkeley, CA 94720, USA}
\affiliation[4]{Department of Physics, Boston University, 590 Commonwealth Avenue, Boston, MA 02215 USA}
\affiliation[5]{Dipartimento di Fisica ``Aldo Pontremoli'', Universit\`a degli Studi di Milano, Via Celoria 16, I-20133 Milano, Italy}
\affiliation[6]{INAF-Osservatorio Astronomico di Brera, Via Brera 28, 20122 Milano, Italy}
\affiliation[7]{Department of Physics \& Astronomy, University College London, Gower Street, London, WC1E 6BT, UK}
\affiliation[8]{Institut d'Estudis Espacials de Catalunya (IEEC), c/ Esteve Terradas 1, Edifici RDIT, Campus PMT-UPC, 08860 Castelldefels, Spain}
\affiliation[9]{Institute of Space Sciences, ICE-CSIC, Campus UAB, Carrer de Can Magrans s/n, 08913 Bellaterra, Barcelona, Spain}
\affiliation[10]{Instituto de F\'{\i}sica, Universidad Nacional Aut\'{o}noma de M\'{e}xico,  Circuito de la Investigaci\'{o}n Cient\'{\i}fica, Ciudad Universitaria, Cd. de M\'{e}xico  C.~P.~04510,  M\'{e}xico}
\affiliation[11]{University of California, Berkeley, 110 Sproul Hall \#5800 Berkeley, CA 94720, USA}
\affiliation[12]{Institut de F\'{i}sica d’Altes Energies (IFAE), The Barcelona Institute of Science and Technology, Edifici Cn, Campus UAB, 08193, Bellaterra (Barcelona), Spain}
\affiliation[13]{Departamento de F\'isica, Universidad de los Andes, Cra. 1 No. 18A-10, Edificio Ip, CP 111711, Bogot\'a, Colombia}
\affiliation[14]{Observatorio Astron\'omico, Universidad de los Andes, Cra. 1 No. 18A-10, Edificio H, CP 111711 Bogot\'a, Colombia}
\affiliation[15]{Institute of Cosmology and Gravitation, University of Portsmouth, Dennis Sciama Building, Portsmouth, PO1 3FX, UK}
\affiliation[16]{Fermi National Accelerator Laboratory, PO Box 500, Batavia, IL 60510, USA}
\affiliation[17]{Steward Observatory, University of Arizona, 933 N. Cherry Avenue, Tucson, AZ 85721, USA}
\affiliation[18]{Center for Cosmology and AstroParticle Physics, The Ohio State University, 191 West Woodruff Avenue, Columbus, OH 43210, USA}
\affiliation[19]{Department of Physics, The Ohio State University, 191 West Woodruff Avenue, Columbus, OH 43210, USA}
\affiliation[20]{The Ohio State University, Columbus, 43210 OH, USA}
\affiliation[21]{Department of Physics, University of Michigan, 450 Church Street, Ann Arbor, MI 48109, USA}
\affiliation[22]{University of Michigan, 500 S. State Street, Ann Arbor, MI 48109, USA}
\affiliation[23]{Department of Physics, The University of Texas at Dallas, 800 W. Campbell Rd., Richardson, TX 75080, USA}
\affiliation[24]{NSF NOIRLab, 950 N. Cherry Ave., Tucson, AZ 85719, USA}
\affiliation[25]{Department of Physics and Astronomy, University of California, Irvine, 92697, USA}
\affiliation[26]{Sorbonne Universit\'{e}, CNRS/IN2P3, Laboratoire de Physique Nucl\'{e}aire et de Hautes Energies (LPNHE), FR-75005 Paris, France}
\affiliation[27]{Departament de F\'{i}sica, Serra H\'{u}nter, Universitat Aut\`{o}noma de Barcelona, 08193 Bellaterra (Barcelona), Spain}
\affiliation[28]{Instituci\'{o} Catalana de Recerca i Estudis Avan\c{c}ats, Passeig de Llu\'{\i}s Companys, 23, 08010 Barcelona, Spain}
\affiliation[29]{Department of Physics and Astronomy, Siena College, 515 Loudon Road, Loudonville, NY 12211, USA}
\affiliation[30]{Department of Physics \& Astronomy and Pittsburgh Particle Physics, Astrophysics, and Cosmology Center (PITT PACC), University of Pittsburgh, 3941 O'Hara Street, Pittsburgh, PA 15260, USA}
\affiliation[31]{IRFU, CEA, Universit\'{e} Paris-Saclay, F-91191 Gif-sur-Yvette, France}
\affiliation[32]{Department of Physics and Astronomy, University of Waterloo, 200 University Ave W, Waterloo, ON N2L 3G1, Canada}
\affiliation[33]{Perimeter Institute for Theoretical Physics, 31 Caroline St. North, Waterloo, ON N2L 2Y5, Canada}
\affiliation[34]{Waterloo Centre for Astrophysics, University of Waterloo, 200 University Ave W, Waterloo, ON N2L 3G1, Canada}
\affiliation[35]{Instituto de Astrof\'{i}sica de Andaluc\'{i}a (CSIC), Glorieta de la Astronom\'{i}a, s/n, E-18008 Granada, Spain}
\affiliation[36]{Departament de F\'isica, EEBE, Universitat Polit\`ecnica de Catalunya, c/Eduard Maristany 10, 08930 Barcelona, Spain}
\affiliation[37]{Department of Physics and Astronomy, Sejong University, 209 Neungdong-ro, Gwangjin-gu, Seoul 05006, Republic of Korea}
\affiliation[38]{CIEMAT, Avenida Complutense 40, E-28040 Madrid, Spain}
\affiliation[39]{Department of Physics \& Astronomy, Ohio University, 139 University Terrace, Athens, OH 45701, USA}
\affiliation[40]{National Astronomical Observatories, Chinese Academy of Sciences, A20 Datun Road, Chaoyang District, Beijing, 100101, P.~R.~China}

\emailAdd{r.cereskaite@sussex.ac.uk}

\abstract{Measurements of galaxy distributions at large cosmic distances capture clustering from the past. In this study, we use a cosmological model to translate these observations into the present-day galaxy distribution. Specifically, we reconstruct the 3D linear matter power spectrum at redshift \( z = 0 \) using Dark Energy Spectroscopic Instrument (DESI) Year 1 (DR1) galaxy clustering data and Cosmic Microwave Background (CMB) observations, assuming the $\Lambda \text{CDM}$ model, and compare it to the result assuming the $w_0w_a \text{CDM}$ model. Building on previous state-of-the-art methods, we apply Effective Field Theory (EFT) modelling of the galaxy power spectrum to account for small-scale effects in the 2-point statistics of galaxy data. Implementation of the EFT approach improves the modelling of the galaxy power spectrum, providing a more robust consistency test of the assumed cosmological model. By casting both CMB and galaxy clustering observations, spanning distinct redshift regimes, into $k$-space, we can identify discrepancies between the datasets of different redshifts, which would indicate potential inaccuracies in the assumed expansion history. While previous studies have shown consistency with \( \Lambda \text{CDM} \), this work extends the analysis with higher-quality data to further test the expansion histories of both \( \Lambda \text{CDM} \) and \( w_0w_a \text{CDM} \). Our findings show that both \( \Lambda \text{CDM} \) and \( w_0w_a \text{CDM} \) provide consistent fits to the linear matter power spectrum recovered from DESI DR1 data.}

\arxivnumber{2507.16590}

\notoc

\begin{document}
\maketitle
\flushbottom

\section{Introduction}\label{sec:Introduction}

Long before the discovery of cosmic acceleration \cite{Riess_1998, Perlmutter_1999}, cosmologists worked to develop models that accurately describe cosmic observations. This discovery, however, marked a turning point, driving the refinement of these models to incorporate the accelerated expansion. The highly successful $\Lambda \text{CDM}$ model emerged as the leading framework, describing a Universe composed of approximately 5\% baryonic matter, 25\% cold dark matter (CDM) and 70\% dark energy (DE) at present time, with less than 1\% contributions from neutrinos and radiation. As the dominant component of the Universe’s energy budget, dark energy is also the sole driver of its accelerated expansion. Analysing cosmic observations at both large and small scales offers valuable insights into the nature of dark energy and its role in shaping the Universe’s evolution.

Recent analyses of the Dark Energy Spectroscopic Instrument (DESI) Year 1 (DR1) and Year 3 (DR2) BAO and Full-Shape spectrum data, in combination with other probes, indicate a notable deviation in expansion history from the standard $\Lambda \rm CDM$ predictions. Using the CPL parametrization \cite{CHEVALLIER_2001, Linder2003}, which allows for a time-varying dark energy (DE) Equation of State (EoS), both data sets suggest a preference for a $w_0 w_a \rm CDM$ cosmology \cite{desicollaboration2024desi1, desicollaboration2024novVII, desicollaboration2025ii}, challenging the standard $\Lambda \rm CDM$ predictions of $w_0 = -1$ and $w_a = 0$. Although the significance of this tension arises primarily from the combination of multiple probes, as individual datasets alone provide limited constraints on dark energy, the same tendency is found in constraints from the SN Ia data \cite{rubin2023union, descollaboration2024dark, Camilleri_2024}. Furthermore, tensions in the values of $H_0$ \cite{Wong_2019, 2020Planck, Freedman_2021, Shah_2021, Riess_2022, Freedman_2020} and potentially $\sigma_8$ ($S_8$) \cite{desicollaboration2024novVII, Abdalla2022, Farren_2024, kids2025} further suggest that the `best-fit' $\Lambda$CDM model cannot fully account for all observations. Together, these discrepancies highlight potential limitations of the $\Lambda$CDM model, motivating extensions that introduce additional parameters and extra degrees of freedom to better align with observations.

Another way to test $\Lambda$CDM is through the diverse observations of the Universe's matter components—whether from the Cosmic Microwave Background (CMB) \cite{2020Planck, Hlozek_2012, camphuis2025spt3gd1cmbtemperature}, Ly-$\alpha$ forest \cite{Chabanier_2019, Lee_2013}, galaxy clustering \cite{desicollaboration2024novVII, Reid_2010, Beutler_2021}, or weak lensing \cite{Chabanier_2019, preston2024reconstructingmatterpowerspectrum, Secco_2022}, all of which trace the underlying matter field. Each observation provides unique insights across different scales and times: the CMB captures the largest scales, originating from the epoch of recombination; galaxy clustering reveals the distribution of dark matter on intermediate scales; weak lensing maps distortions in the shapes of background galaxies caused by the gravitational influence of foreground matter on intermediate to smaller scales; and the Ly-$\alpha$ forest traces the neutral hydrogen distribution in the intergalactic medium, covering the smallest scales among these tracers. Together, these measurements build a cohesive picture of the matter power spectrum across a vast range of scales offering a robust test of the cosmological model. However, while these observables trace the matter field in different ways, they do not directly measure the matter power spectrum. Galaxy clustering is influenced by complex astrophysical processes, including how galaxies are distributed within dark matter halos, which affects their relationship with the underlying matter field, while CMB anisotropies primarily reflect the conditions of the early universe. To understand the large-scale structure as it is today and to test the consistency of the cosmological model across cosmic time, we need to construct the underlying matter power spectrum from these observations and extrapolate to the present epoch (i.e., $z = 0$). This provides a consistent basis for evaluating the evolution of cosmic structure across different redshifts and across different cosmological probes, effectively disentangling late and early time physics.

Recent years have seen several studies demonstrating that reconstructing the matter power spectrum from observations is a valuable approach for probing cosmology and potentially resolving existing tensions \cite{Mu_oz_2020, preston2024reconstructingmatterpowerspectrum}. Numerous works have demonstrated that the recovered spectra from various datasets align well with the theoretical predictions of the linear matter power spectrum in the $\Lambda$CDM model \cite{Chabanier_2019, Reid_2010, preston2024reconstructingmatterpowerspectrum, Tegmark_2002}. For example, analyses using UV galaxy luminosity function (LF) data from the Hubble Space Telescope (HST) \cite{Sabti_2022}, CMB observations from the Atacama Cosmology Telescope (ACT) \cite{Hlozek_2012}, dwarf galaxy kinematic and structure data \cite{esteban2024milkywaysatellitevelocities} and SDSS galaxy clustering data \cite{Reid_2010} have shown strong agreement with expectations from $\Lambda$CDM. However, a recent analysis of eBOSS Ly-$\alpha$ forest data presents a significant exception, reporting a 5$\sigma$ tension with CMB-based inferences of the linear matter power spectrum \cite{rogers20245sigmatensionplanck}. This confirms a discrepancy first identified at the 3$\sigma$ level in earlier work \cite{Palanque_Delabrouille_2020}, highlighting the need for further investigation.

While methods for modelling the galaxy power spectrum have been extensively studied and developed, comparatively little attention has been given to applying these modelling methods in reverse — that is, to extract the matter power spectrum from galaxy clustering data itself. The state of the art for estimating the linear matter power spectrum from galaxy clustering data is described in \cite{Chabanier_2019} and in more detail in \cite{Reid_2010}. Both approaches assume a parameterized function that accounts for galaxy bias — the statistical discrepancy between the observed galaxy distribution and the underlying matter distribution due to the way galaxies form and cluster — and use it to compute the linear matter power spectrum which describes the distribution of matter fluctuations as predicted by linear perturbation theory, before nonlinear structure formation becomes significant. 

With the advent of Stage IV probes, such as Euclid \cite{Euclid2020}, LSST \cite{LSST2009}, Roman \cite{Roman2015} and Square Kilometre Array (SKA) \cite{SKA2020} providing more detailed measurements of the large-scale distribution of galaxies, existing methods to account for anisotropic effects, such as the Alcock-Paczynski (AP) effect \cite{Alcock1979, Padmanabhan_2008, Matsubara_1996} and Redshift Space Distortions (RSD) \cite{Pellejero_Iba_ez_2022, Beutler_2019}, need to be upgraded to match the improved data quality and the increasing complexity of modern surveys. Recent advances in the Effective Field Theory of Large Scale Structure (EFTofLSS) \cite{ivanov2022effectivefieldtheorylarge, Carrasco2012, Chen_2020} have provided the tools to model these effects more accurately. EFTofLSS extends perturbation theory by introducing corrections that account for small-scale nonlinear dynamics, while maintaining predictive power at large scales. It incorporates effective parameters, or counterterms, to encapsulate the influence of unresolved small-scale physics, and employs renormalization techniques to ensure that these effects are consistently accounted for in large-scale predictions.

\begin{figure}
\centering
    \includegraphics[width=\textwidth]{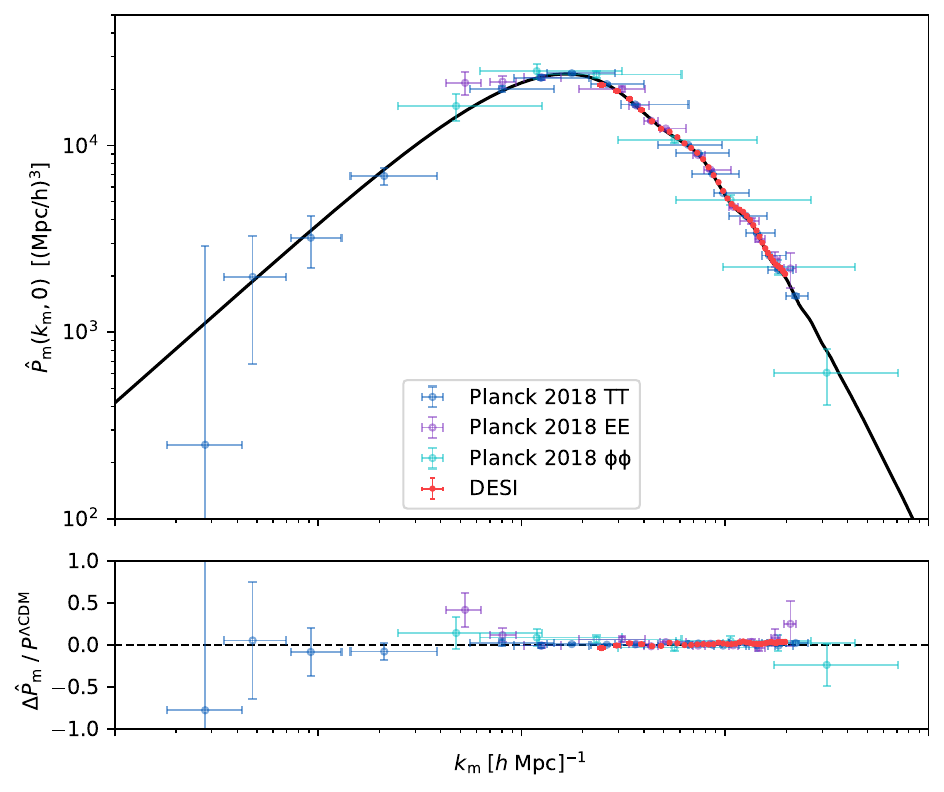}
    \caption{\textit{Top:} Inference of the 3D linear matter power spectrum at \( z = 0 \) under the \( \Lambda \text{CDM} \) model. The data points, spanning from large to intermediate scales, represent CMB data from Planck (shown in blue, purple, and teal). Intermediate scale data from DESI Y1 is indicated by red points. \textit{Bottom:} Residual plot showing deviations of CMB and galaxy clustering data points from Planck’s best-fit $\Lambda$CDM 3D linear matter power spectrum. The dashed line indicates zero deviation. All vertical errorbars represent 1$\sigma$ uncertainties. See Table \ref{tab:goodness-of-fit-table} for $\chi^2$ values.}
    \label{fig:final-plot-desi-all}
\end{figure}

The Dark Energy Spectroscopic Instrument (DESI) is the first operational Stage IV \cite{DE_taskforce} galaxy survey with its primary objectives being to measure the expansion history of the Universe and the growth rate of large-scale structure (LSS) \cite{levi2013desiexperimentwhitepapersnowmass, desicollaboration2016desiexperimentisciencetargeting}. Over a five-year survey spanning 14,200 square degrees, DESI is collecting spectra for approximately 40 million galaxies and quasars \cite{Guy_2023}. The survey targets five distinct tracers: low-redshift galaxies from the Bright Galaxy Survey (BGS) \cite{Hahn_2023}, Luminous Red Galaxies (LRG) \cite{Zhou_2023}, Emission Line Galaxies (ELG) \cite{Raichoor_2023}, Quasars (QSO) \cite{Chaussidon_2023}, and the Ly-$\alpha$ forest features in quasar spectra \cite{Myers_2023}. In this work, we focus on reconstructing the linear matter power spectrum from the Full-Shape data of these galaxy tracers, except for Ly-$\alpha$ forest, as the data for this tracer was not fully available at the time of writing this paper. Full-Shape data refers to the detailed information captured from the full galaxy clustering signal, including both the shape and BAO positions of the redshift-space power spectrum, rather than relying solely on geometrical information from the BAO peaks. We begin with a concise description of the datasets used, including CMB and DESI's Y1 galaxy power spectrum data. We then explain the methodology and the improvements introduced in the analyses of these datasets, building on the work done by \cite{Chabanier_2019}. Finally, we present our reconstructions of the linear matter power spectrum from the data, the combined results of which are presented in Figure~\ref{fig:final-plot-desi-all}, before concluding with a discussion of potential extensions and improvements for future work.

\section{Data}\label{sec:Data}

\subsection*{Cosmic Microwave Background}

The analysis of CMB data in this work follows the approach outlined in \cite{Chabanier_2019} using the same dataset. Specifically, the Planck 2018 temperature, polarization and lensing reconstructed power spectra were used for CMB datapoints. For spherical harmonic modes $\ell > 30$, rebinned bandpowers and covariance matrices were used from the $Plik$ likelihood \cite{PlikLike2020}. For $\ell < 30$, only temperature $C_\ell$ values from the $Commander$ likelihood \cite{PlanckCommander2020} were used, with polarization and lensing data excluded due to their sensitivity to reionization models, as mentioned in \cite{Chabanier_2019} and references therein. As the focus of this study was on improvements in galaxy clustering, the CMB datasets and methods remained unchanged from \cite{Chabanier_2019}, which based their analysis on \cite{Tegmark_2002}.

\subsection*{Galaxy Clustering}
The analysis of galaxy clustering data uses the DESI galaxy and quasar sample from DR1, replacing the Sloan Digital Sky Survey (SDSS) seventh data release \cite{Reid_2010} used previously in \cite{Chabanier_2019}. In particular, we use 2-point clustering statistics in Fourier space, described extensively in \cite{desicollaboration2024novII} and the corresponding EZmock-based covariance matrices, described in \cite{desicollaboration2024novVII,ross2024,Forero-Sanchez:2024bjh,Chuang_2014}. While \cite{Chabanier_2019} relied solely on measurements from Luminous Red Galaxies (LRGs) \cite{Chabanier_2019}, our analysis incorporates all DESI tracers --- namely Bright Galaxy Sample (BGS), Emission Line Galaxies (ELG) and quasars (QSO), as well as the LRGs --- to obtain the final plot. However, we exclude the ELG1 bin ($0.8 < z < 1.1$) due to residual uncorrected large-scale imaging systematics, as described in \cite{desicollaboration2025desi2024vfullshape}. We also recover the matter power spectrum using individual tracers and assess their compatibility.

\section{Methodology}\label{sec:Methodology}

The prevailing methodology for recovering the underlying $z=0$ matter power spectrum from high-redshift observations is thoroughly detailed in \cite{Tegmark_2002}, which outlines a process for recovering $P_{\rm m} (k)$ when the transfer functions are known. The transfer functions describe how perturbations in the early universe evolve into the matter distribution we observe today. This method compares rebinned data from different observations, including the CMB and galaxy clustering, with theoretical predictions of the corresponding power spectrum by taking the ratio of data over theory, where the theory is expressed as integrals over linear scales $k$. These integrals
incorporate the survey window function and its convolution with the power spectrum, which is computed based on a given cosmological model. The data points are plotted at the $k$-value corresponding to the median of the distribution arising from rebinning, with horizontal bars spanning from the 20th to the 80th percentile. Later on, \cite{Chabanier_2019} applies this methodology for data points representing the CMB, while they recover galaxy clustering data points following a simple division by an assumed galaxy bias function. In our work, we follow the same general approach for the CMB analysis but improve upon the treatment of galaxy clustering data by refining the methodology to more accurately account for scale-dependent effects and observational systematics (see \ref{sec:Methodology-CMB} and \ref{sec:Methodology-GC} for details).

\subsection{Cosmic Microwave Background (CMB)}\label{sec:Methodology-CMB}

\begin{figure}
\centering
    \includegraphics[width=\textwidth]{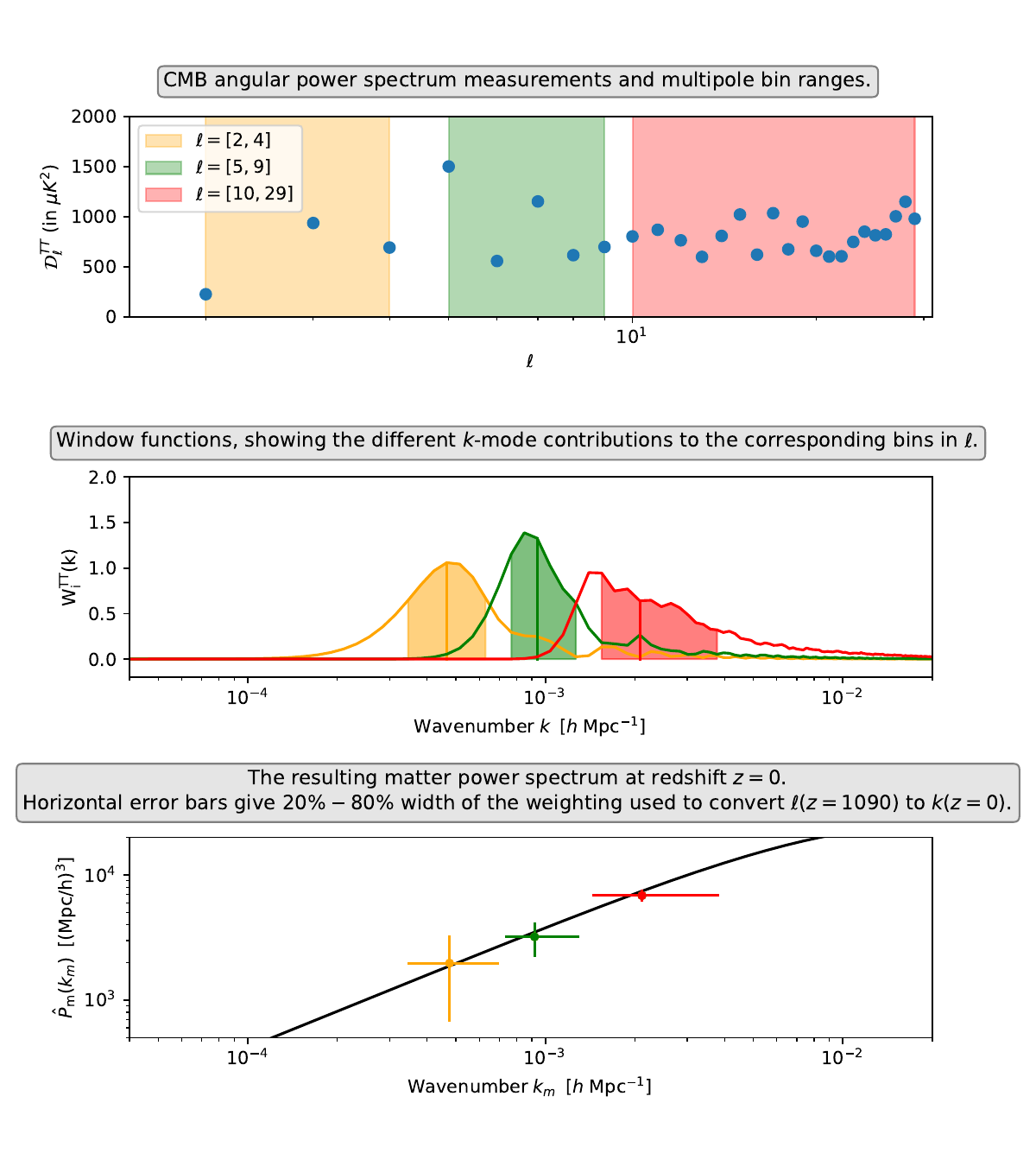}
    \caption{Schematic illustrating the mapping of CMB angular power spectrum measurements to Fourier space. The top panel displays measurements in the low-$\ell$ regime, which are binned in $\ell$ as indicated by the shaded regions. The middle panel corresponds to the windowed transfer functions of CMB, with each curve representing a contribution of $k$-modes to the $\ell$ bin. These are treated as probability density functions to extract the values of $k$, at which the CMB datapoint is plotted. The horizontal error bars represent the $20–80\%$ range of the $k$ contribution to each specific multipole bin, while the vertical bold line in the middle panel marks the median of this distribution, $k_{m,i}$ , i.e. the scale below which $50\%$ of the contribution lies. This median value is used as the relevant scale for inference of the linear matter power spectrum in the bottom panel, as described in Equation~\ref{eq:1} and the text thereafter.}
    \label{fig:CMB-explanation}
\end{figure}

This analysis involves extrapolating high-redshift observational data to low redshift using an assumed cosmological model. The original high-redshift data show how much power is present across a range of angular scales. However, there is no direct one-to-one mapping between angular scales ($\ell$) at high redshift and Fourier-space scales ($k$) at low redshift, as a range of high-redshift angular scales contribute to each low-redshift scale. To account for this, we perform a convolution --- a weighted averaging of the CMB data at different $\ell$-modes --- to determine the corresponding low-redshift matter power spectrum. This process is illustrated in Figure \ref{fig:CMB-explanation}: the top panel displays the input CMB power spectrum, the middle panel shows the weighting applied to each high-redshift $\ell$-bin for conversion to a low-redshift $k$-bin, and the bottom panel presents the resulting power spectrum at $z=0$. 

We keep the established approach for the CMB, which not only allows us to compare our Python code with the Julia code \texttt{mpk\_compilation} provided by \cite{Chabanier_2019}, but also serves as a guide when developing functions for computing $k_{m,i}$ values for galaxy clustering data points.\footnote{https://github.com/marius311/mpk\_compilation.} The resulting data points, denoted as $\hat{P}_{\rm m} (k_{m,i}, 0)$, are obtained by:
\begin{equation}
\label{eq:1}
    \begin{aligned}
        \hat{P}_{\rm m} (k_{m,i}, 0) =&  {P^{\Lambda \text{CDM}} (k_{m,i}, 0)} \frac{ \mathcal{C}_i }{\mathcal{C}_{i}^{\Lambda \text{CDM}}}  \\ 
        =& {P^{\Lambda\text{CDM}}(k_{m,i}, 0)} \frac{\sum_{\ell} \textbf{W}_{i \ell}\mathcal{C_{\ell}}} {\int \sum_{\ell} \textbf{W}_{i \ell} W_{\ell}(k) \Delta_\mathcal{R}^2(k)\, \,d \ln k} \\ 
        =& {P^{\Lambda\text{CDM}}(k_{m,i}, 0)} \frac{\sum_{\ell} \textbf{W}_{i \ell}\mathcal{C_{\ell}}} {\int \underbrace{\textnormal{W}_{i} (k)}_{\textnormal{CMB Kernels}} \Delta_{\mathcal{R}}^{2}(k) \, \,d \ln k}.
    \end{aligned}
\end{equation}

\noindent Here, $P^{\Lambda\text{CDM}}(k_{m,i}, 0)$ is the theoretical linear matter power spectrum at $z=0$ assuming $\Lambda \rm CDM$ cosmology, $\Delta_{\mathcal{R}}^2 (k)$ is the dimensionless primordial power spectrum of curvature perturbations and $k_{m,i}$ represents the median value of $k$, at which the recovered matter power spectrum data point $\hat{P}_{\rm m}(k_{m,i}, 0)$ is plotted. The characteristic scale $k_{m,i}$ for bandpower $i$ is obtained by combining the CMB multipoles through the survey’s window matrix $\mathbf{W}_{i{\ell}}$, which projects theoretical predictions into the same bins as the data. This defines a normalized kernel $\Pi_i(k)$ describing the relative weight of each Fourier mode $k$ to the $i^{\rm th}$ bandpower. The $k_{m,i}$ is then the median value of this distribution, i.e. the scale below which half of the total contribution arises (Equation \ref{eq:3.2}).

\begin{equation}
\int_{k_{\min}}^{k_{m,i}} \Pi_i(k)\, d\ln k = 0.5,
\qquad \text{with} \quad
\Pi_i(k) = \frac{\textnormal{W}_i(k)}{\int_{k_{\min}}^{k_{\max}} \textnormal{W}_i(k')\, d\ln k'}.
\label{eq:3.2}
\end{equation}

In this way, the matrix-weighted transfer function $\textnormal{W}_{i}(k)$ describes how different Fourier modes $k$ contribute to the observed CMB power spectrum bandpowers, as pictured in Figure \ref{fig:CMB-explanation}. The shaded region in the middle panel corresponds to kernel values spanning from 20\% to 80\%, denoted by the edges of the shaded area. The bold middle line corresponds to 50\% --- the median value of $k$ of the distribution ($k_{m,i}$). These quantities define the representative wavenumbers at which the matter power spectrum is inferred from the data.

The rebinning of CMB transfer window functions accounts for the fact that our measurements are constrained by incomplete sky coverage, meaning the measurement represents a compressed version of the CMB bandpowers, expressed as the expectation value of the angular power spectrum $ \mathcal{C}_i $. $\mathcal{C}_i^{\Lambda \text{CDM}}$ is the theoretically computed angular power spectrum convolved with the survey's window matrix under the assumption of a $\Lambda$CDM background. Expressing $\mathcal{C}_i^{\Lambda \text{CDM}}$ as an integral over scales, where the primordial power spectrum is convolved with the weighted transfer function $\textnormal{W}_{i}(k)$ leads to the second equality. This ratio of the data to the constructed model that replicates the data can be interpreted as a scaling factor, which in turn increases or decreases the computed $z=0$ matter power spectrum at scales $k_{m,i}$. Equation \ref{eq:1} also assumes linear structure growth, as computing transfer functions in the case of CMB is not trivial — it requires taking into account neutrinos and other factors such as dark matter, baryon-photon interactions and the evolution of the universe through various phases (e.g., radiation domination, matter domination, and dark energy domination).

\subsection{Galaxy Clustering}\label{sec:Methodology-GC}

For the treatment of the galaxy power spectrum, a similar approach was adopted as in \cite{Tegmark_2002}. However, due to the improved precision of DESI's observations, previously adopted models are insufficient for addressing the small-scale effects and redshift space distortions arising from the  peculiar velocities of galaxies, which must be considered in the analysis of galaxy data. As peculiar velocities of galaxies contain information about the growth of Large Scale Structure (LSS), they lead to anisotropies in the observations of galaxy clustering signal. More recently,  Effective Field Theory (EFT) models were introduced for accurate modelling of the galaxy power spectrum in both real and redshift space \cite{Chen_2020, Chen_2021}. EFT improves upon earlier techniques by systematically incorporating corrections for small-scale nonlinearities, allowing for more accurate predictions of the galaxy clustering signal, particularly in regions where these nonlinear effects begin to manifest and influence the observations. In this work, we focus on \texttt{velocileptors} \cite{Chen_2020, Chen_2021}, a Python code designed to compute the one-loop velocity statistics and redshift-space power spectrum using both Eulerian Perturbation Theory (EPT) and Lagrangian Perturbation Theory (LPT).\footnote{https://github.com/sfschen/velocileptors/.} Our analysis follows the official DESI likelihood pipeline, \texttt{desilike}, including the adopted prior ranges and other analysis choices, as described in \cite{desicollaboration2025desi2024vfullshape}.\footnote{https://github.com/cosmodesi/desilike/.}

Within the EFT-based LPT framework of \texttt{velocileptors}, the galaxy power spectrum is given by:
\begin{equation}
\label{eq:2}
    P_{g}^{\rm EFT}(\textbf{k}) = P_{\textnormal{LPT}} (\textbf{k}) + k^2(\alpha_0 + \alpha_2 \mu^2 + \alpha_4 \mu^4) P_{\textnormal{Zel}}(\textbf{k}) + (\textnormal{SN}_0 + \textnormal{SN}_2 k^2 \mu^2 + \textnormal{SN}_4 k^4 \mu^4).
\end{equation}
\noindent Here $\mu$ is the cosine of the angle between the line-of-sight and the wavevector $\textbf{k}$, with $k$ denoting its magnitude. $P_{\textnormal{Zel}}(\textbf{k})$ refers to the linear Zeldovich approximation, $\textnormal{SN}_0, \textnormal{SN}_2, \textnormal{SN}_4$ are stochastic terms accounting for residual shot noise and other uncorrelated small-scale effects and $\alpha_0, \alpha_2, \alpha_4$ are counterterms to account for the effects of small-scale physics below the EFT cut-off scale (for more details, see \cite{Chen_2020, Chen_2021}). Galaxy power spectrum multipoles are then given by:
\begin{equation}
\label{eq:3}
    P_{\ell} (k) = \frac{(2\ell + 1)}{2} \int_{-1}^1 P_g(k, \mu) \mathcal{L}_{\ell} (\mu) \,d\mu.
\end{equation}
However, with the transition from the previously used linear bias model to EFT-based models there are two challenges to consider: first, the procedure of extracting $k_{m,i}$ values, as it was done in the case of CMB, is now complicated by the necessity to separate out matter power spectrum from EFT counterterms --- a procedure that is rather laborious. Furthermore, the second challenge is DESI's improved treatment of the window function \cite{pinon2024}, which introduces an additional level of complexity to the accurate  extraction of $k_{m,i}$ values. Therefore, we propose a simplified approach for extraction of $k_{m,i}$ values for galaxy clustering data using a Kaiser-model approximation.

\noindent We begin by relating the galaxy power spectrum to the matter power spectrum at a given redshift, assuming the linear Kaiser model for this simplified presentation:
\begin{equation}
\label{eq:4}
    P_{g}^{\rm K}(k, \mu) = b^2 (1 + \beta \mu^2)^2 P^{\Lambda \text{CDM}} (k).
\end{equation}
\noindent The quantity $b$ represents the galaxy-matter linear bias parameter, while $\beta$ is defined as $\beta(z) = f(z)/b$, where $f(z)$ is the linear growth rate. The growth rate $f (z)$ is given by:
\begin{equation}
    f(z)= \frac{d \ln D(z)}{d \ln a},
\end{equation}
\noindent where $D(z)$ is the linear growth function, describing how density fluctuations evolve over time, and $a$ is the scale factor.

\noindent We define the window-convolved galaxy power spectrum multipoles as:
\begin{equation}
\label{eq:5}
    P_{\ell}^{\textnormal{conv}} (k,z) = \int k'^2 \sum_{\ell'} W_{\ell \ell'}(k,k') P_{\ell'}(k',z) \,d k',
\end{equation}
\noindent where $W_{\ell \ell'}(k,k')$ is the window function expanded into the Legendre multipole space describing the mixing of modes due to the survey geometry and footprint on the sky. 
Substituting Equation~\ref{eq:3} as $P_{\ell'}(k', z)$ and Eq.~\ref{eq:4} as $P_g(k',\mu)$ into the definition of window-convolved galaxy power spectrum multipoles in Equation~\ref{eq:5} we arrive to the final expression of $P_{\ell}^{\textnormal{conv}} (k, z)$:
\begin{equation}
\label{eq:6}
    P_{\ell}^{\textnormal{conv}} (k, z) = \int P^{\Lambda \text{CDM}} (k', z) 
    \underbrace{k'^2 \sum_{\ell'} W_{\ell \ell'}(k,k') \left[ \frac{(2\ell' + 1)}{2} \int_{-1}^1 b^2 (1+\beta \mu^2)^2 \mathcal{L}_{\ell'} (\mu) \,d\mu \right]}_{\text{`Galaxy Kernels'}} \,d k',
\end{equation}
\noindent where we now have a form for the distribution function, denoted as `Galaxy Kernels', which can be used to find $k_{m,i}$. This term represents the mixing of multipoles at different $k$ values and a combined contribution of different multipoles $\ell'$ to the total power spectrum at a given $k$. The inferred matter power spectrum from galaxy Full-Shape multipoles at the effective redshift of the tracer bin $\hat{P}_{\rm m, \ell} (k_{m,i}, z)$ is obtained by:

\begin{equation}
\label{eq:7}
     \hat{P}_{\rm m, \ell} (k_{m,i}, z) = P^{\Lambda \textnormal{CDM}}(k_{m,i}, z) \frac{ P_{\ell}^\textnormal{conv, obs}(k_{{\rm eff}}, z)} {P_{\ell}^\textnormal{conv}(k_{{\rm eff}}, z)}.
\end{equation}

\noindent Here $P_{\ell}^\textnormal{conv, obs}(k_{\rm eff}, z)$ is the observed galaxy clustering multipole data, binned at $k_{\rm eff}$.

\begin{figure}
\centering
    \includegraphics[width=\textwidth]{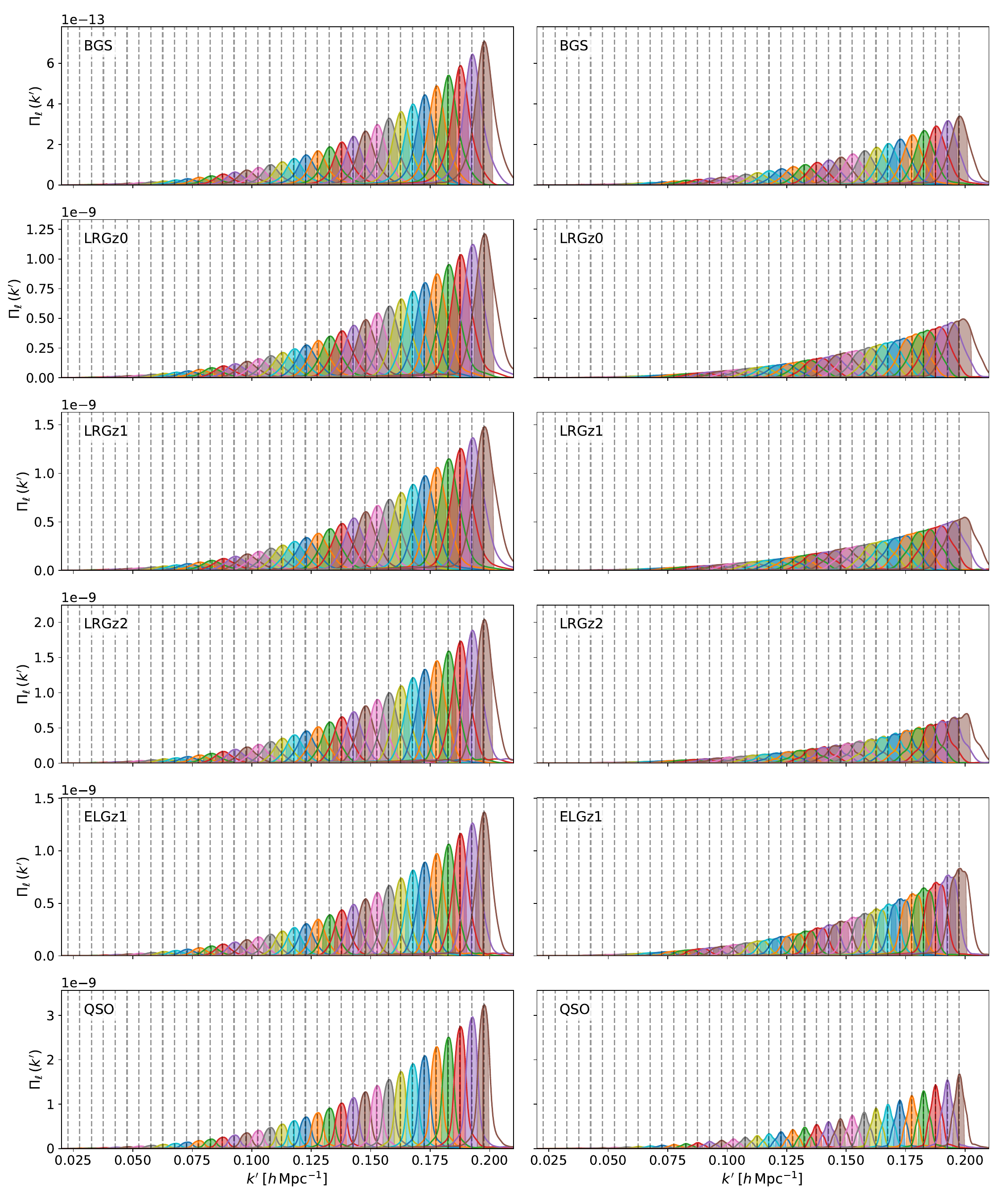}
    \caption{Galaxy Kernels of all tracers extracted for monopole (left column) and quadrupole (right column). The galaxy kernels, defined in Equation~\ref{eq:6} under the Kaiser approximation, are used to compute the median $k_{m,i}$} values for application in Equation~\ref{eq:7}. The vertical dashed lines correspond to $k_{\textnormal{eff}, i}$ values.
    \label{fig:galaxy-kernels}
\end{figure}

The `Galaxy Kernels' and their corresponding $k_{m,i}$ results are shown in Figure~\ref{fig:galaxy-kernels}. These kernels are employed analogously to the CMB case, by substituting the galaxy kernel into Equation~\ref{eq:3.2} in place of the CMB kernel. The linear Kaiser bias values entering the determination of $k_{m,i}$ are obtained from the $\chi^2$ minimization procedure described in text below and are $b = 1.611, 1.840, 2.035, 2.181, 1.408,$ and $2.262$ for BGS, LRG1, LRG2, LRG3, ELG2, and QSO, respectively. We note that the resulting $k_{m,i}$ values differ from the effective scales $k_{\text{eff}, i}$ obtained from the \texttt{desilike} likelihood pipeline, which directly handles window convolution with the theory and incorporates all necessary corrections internally. We compared the extracted $k_{m,i}$ values to the corresponding $k_{\text{eff}, i}$ values and found differences of up to 18\% for the monopole and up to 21\% for the quadrupole, with the largest deviations appearing for the LRG2 ($0.6 \leq z \leq 0.8$) tracer on the largest scales. These discrepancies are most prominent at large scales, while on small scales the difference between $k_{m,i}$ and $k_{\text{eff}, i}$ is consistently below the percent level across all tracers.

We hereafter present and discuss the conceptual differences between $k_{\text{eff}, i}$ and $k_{m,i}$. The effective scale, $k_{\text{eff}, i}$, is defined as the spherical average of the wavenumbers $k$ of the Fourier modes contributing to the measurement, weighted by the number of modes available in a given bin. Fundamentally, $k_{m,i}$ and $k_{\textnormal{eff}, i}$ are two different quantities --- in the case of the CMB, $k_{m,i}$ is essential because the measurement is made in angular space, and each multipole moment receives contributions from a broad range of $k$-modes. In contrast, the galaxy power spectrum from DESI is computed in Fourier space, where modes are grouped into bins defined over discrete ranges of $k$. Here, $k_{\textnormal{eff}, i}$ represents the weighted average of the $k$-values within each bin and is therefore a more appropriate choice for characterizing the data. To be consistent with the analysis of CMB, we define the bin centres in our plots using $k_{m,i}$ with errorbars corresponding to the range of $k$-modes contributing to each bin. The final expression for the inferred matter power spectrum from galaxy clustering follows the form of Equation \ref{eq:7}. This formulation allows for a direct comparison between the galaxy power data points at the $k_{\text{eff}, i}$ values and the theoretical estimate of galaxy power spectrum multipoles, which are convolved with the survey window and account for all related effects at the same $k_{\text{eff}, i}$ values. The ratio of the two quantities scales the matter power spectrum, computed at the given $k_{m,i}$ values and at the effective redshift of the data for the fiducial cosmology being tested. The use of Kaiser modeling in determining $k_{m,i}$ offers a conservative approximation to characterising the galaxy bandpowers at their corresponding bins $k_{m,i}$ and presents a simpler modelling framework, which accounts for wide and conservative range of scales contributing to the measurement, avoiding the need to separate biases and stochastic terms when transitioning to more complex EFT models.

We also validate our choice of $k$-range in Appendix \ref{appendix}, where we show the recovered matter power spectrum fits the linear prediction when applying our methodology to DESI's BGS mock catalogues with known underlying cosmology. Any potential deviations from linear power spectrum in the quasi-linear region would be the most dominant in this low-redshift tracer, which proves that the method is robust in our chosen ranges of $k$.

The resulting $\hat{P}_{\rm m,{\ell}}(k_{m,i}, z)$ from galaxy clustering data can then be translated to the primordial power spectrum or projected to redshift $z = 0$ using the transfer functions, which are computed using a Boltzmann code assuming a fiducial $\Lambda$CDM background:
\begin{equation}
\label{eq:9}
     \hat{P}_{\rm m,{\ell}}(k_{m,i}, 0) = \frac{T^2(k_{m,i}, 0)}{T^2(k_{m,i}, z)} \hat{P}_{\rm m,\ell}(k_{m,i}, z).
\end{equation}
The galaxy methodology can be broken down into the following steps: first, all cosmological parameters are fixed to DESI’s fiducial $\Lambda$CDM values. The EFT model \texttt{velocileptors} is used to fit the data. The nuisance parameters of the model are optimized by minimizing the $\chi^2$ (or, equivalently, maximizing the likelihood) with respect to these parameters, yielding their best-fit values under the fixed cosmology. These best-fit nuisance parameters are then plugged into the denominator of Equation~\ref{eq:7} to produce a theoretical prediction of the power spectrum multipoles. An inference of the matter power spectrum extrapolated to present times is achieved using transfer functions computed by assuming the same fiducial cosmology following Equation~\ref{eq:9}. 

We use the generalized least-squares estimate when combining the inferred matter power spectrum datapoints from monopole and quadrupole of each tracer:
\begin{equation}
    \mathbf{\bar{P}}_{\rm m} = \left( \mathbf{A}^\mathrm{T}        \mathbf{C}^{-1} \mathbf{A} \right)^{-1} 
                   \mathbf{A}^\mathrm{T} \mathbf{C}^{-1} \mathbf{\hat{P}}_{\rm m},
\end{equation}
\noindent where $\mathbf{\hat{P}}_{\rm m}$ is a data vector of present-day matter power spectrum as inferred from galaxy power spectrum multipoles. $\mathbf{C}$ is the $2N \times 2N$ covariance matrix, describing the statistical uncertainties and correlations of measured galaxy power spectrum multipoles, that is rescaled by applying data-to-theory ratio at the effective redshift and the transfer function corrections to the monopole and quadrupole components. $\mathbf{A}$ is a $2N \times N$ mapping matrix that relates the underlying $k$-bin values to the measured monopole and quadrupole, constructed by vertically stacking two $N\times N$ identity matrices and $N$ is the number of $k$-bins per multipole. Finally, $\mathbf{\bar{P}}_{\rm m}$ represents the minimum-variance combined estimate of the underlying matter power spectrum at each $k$-bin. Note that the covariance for this estimate corresponds to $\mathrm{Cov}(\mathbf{\bar{P}}_{\rm m}) = (\mathbf{A}^\mathrm{T} \mathbf{C}^{-1} \mathbf{A})^{-1}$. To combine measurements from multiple tracers---either from combination of monopole and quadrupole or treating multipoles separately---we compute a weighted average of the inferred matter power spectra. Each tracer is associated with a covariance matrix that encodes statistical uncertainties and correlations; if $k$ errors are included, these are propagated into the covariance as well. The final, combined measurement at each $k_{m,i}$, shown in Figure~\ref{fig:final-plot-desi-all}, is obtained by inverse-covariance weighting across all tracers:

\begin{equation}
\bar{\mathbf{P}}_{\rm m}^\mathrm{all} = \mathbf{C}^\mathrm{all} 
\sum_{i=1}^{N_\mathrm{tracers}} \mathbf{C}_i^{-1} \, \bar{\mathbf{P}}_{\text{m},i}, 
\quad
\mathbf{C}^\mathrm{all} = \left( \sum_{i=1}^{N_\mathrm{tracers}} \mathbf{C}_i^{-1} \right)^{-1}.
\end{equation}

\noindent Here, $\mathbf{\bar{P}}_{\rm m}^{\rm all}$ denotes the combined matter power spectrum from all tracers. The sum runs over number of tracers $N_{\rm tracers}$, with $\mathbf{\bar{P}}_{\text{m},i}$ representing the measurement for tracer $i$, and $\mathbf{C}_i$ its associated covariance matrix. The matrix $\mathbf{C}^{\rm all}$ is the covariance of the combined measurement. $\mathbf{\bar{P}}_{\rm m}^{\rm all}$ provides a representative measurement of matter power spectrum from all of DESI’s tracers and galaxy redshift bins, under the assumption that their non-overlapping redshift coverage allows them to be treated as independent.

Recent cosmological results from DESI Y1 BAO and Full-Shape data suggest a preference for a time-varying equation of state for Dark Energy. To explore this possibility, we took an additional step and re-applied our analysis in the context of the $w_0w_a\text{CDM}$ cosmology. For this, we fixed the cosmological parameters to the marginalized mean values obtained in \cite{desicollaboration2024novVII}, specifically using the result with $\Omega_m = 0.3142 \pm 0.0063$, $\sigma_8 = 0.8163 \pm 0.0083$, $S_8 = 0.8353 \pm 0.0092$, $H_0 = 67.48 \pm 0.62$, $w_0 = -0.761 \pm 0.065$, $w_a = -0.96^{+0.30}_{-0.26}$. These values were obtained through a Bayesian analysis combining DESI DR1 BAO and Full-Shape data with CMB constraints from Planck PR3 \cite{2020Planck}, Planck PR4 \cite{Carron_2022} CMB lensing reconstruction, ACT DR6 \cite{Madhavacheril_2024}, and supernova data from DES-SN5YR \cite{DES-SN5YR}. For this updated cosmology, we recompute the transfer functions and refit the nuisance parameters of the EFT model to obtain best-fit values. While we find some small shifts in these parameters relative to the $\Lambda$CDM case, they remain well within the statistical uncertainties, indicating minimal sensitivity of the nuisance calibration to the change in cosmology (see Section \ref{sec:Results} for details).

For the theoretical cosmological calculations in this work, we employed the Boltzmann code \texttt{CAMB} \cite{Lewis_2000}.\footnote{https://github.com/cmbant/CAMB.} The best-fit values for nuisance parameters were obtained by employing  \texttt{iminuit} \cite{iminuit} within the \texttt{desilike} framework, which is a comprehensive tool that encompasses all necessary methodologies for DESI likelihood analysis, including the implementation of \texttt{velocileptors}.\footnote{https://github.com/scikit-hep/iminuit.}


\section{Results}\label{sec:Results}

\begin{figure}
\centering
    \includegraphics[width=\textwidth]{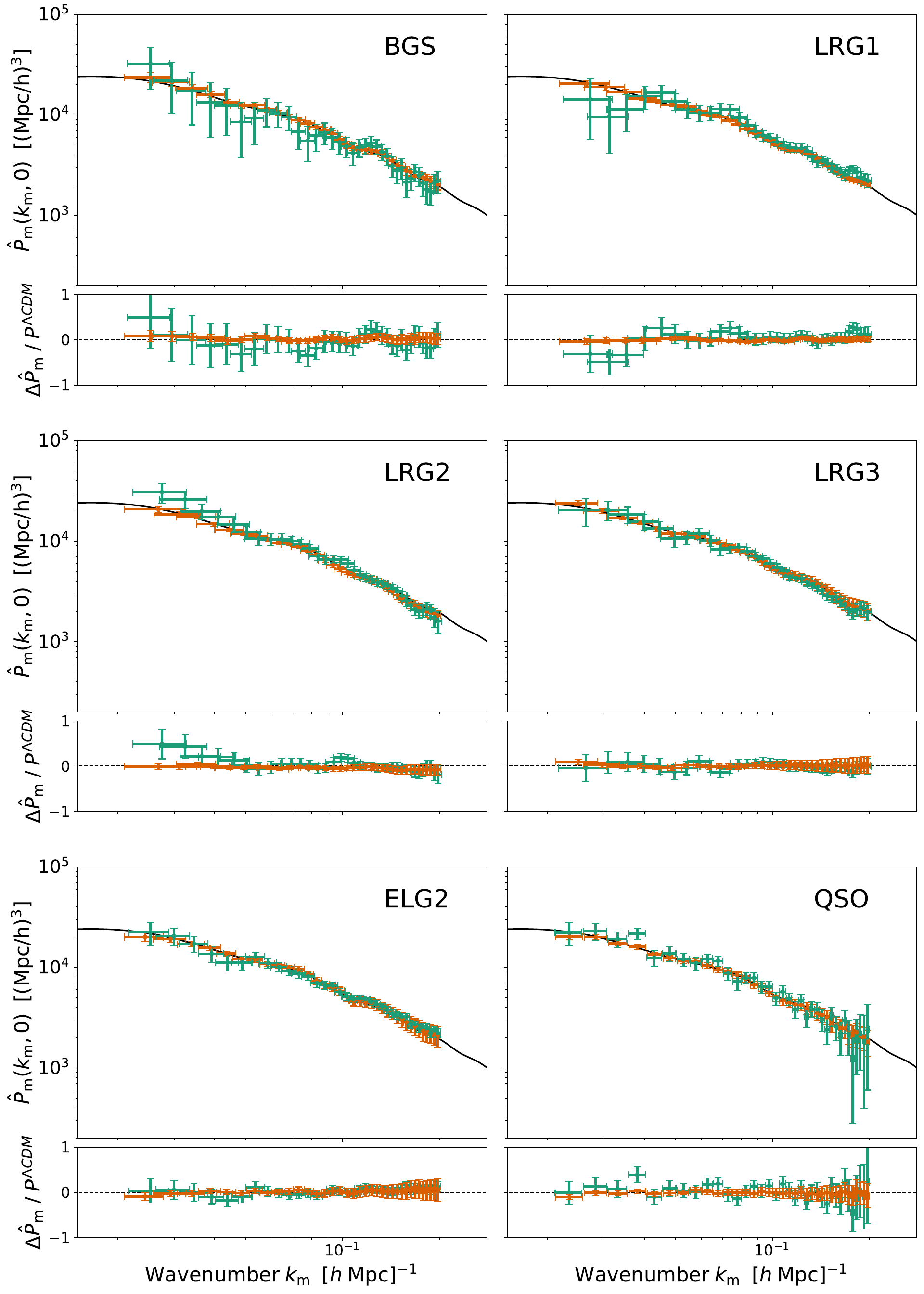}
    \caption{The 3D linear matter power spectrum at $z = 0$, estimated separately for the monopole (orange) and quadrupole (green) measurements from each galaxy clustering tracer in DESI Y1 data, with each data point evaluated at its corresponding $k_{m,i}$ value.}
    \label{fig:all-tracers-combined}
\end{figure}

The best-fit values of the nuisance parameters, computed using the $\chi^2$ minimization method, are presented in Table~\ref{tab:bestfit-params}. These values are then used to compute the theoretical galaxy power spectrum within the \texttt{velocileptors} framework, providing a reference model for the data-to-theory ratio in Equation~\ref{eq:7}. The resulting matter power spectrum, extrapolated to the present time and obtained by averaging the monopole and quadrupole of the DESI Y1 Full-Shape data from all tracers, is shown in Figure \ref{fig:final-plot-desi-all}. This is based on the \(\Lambda\)CDM cosmological model and EFT modelling of the galaxy power spectrum. The matter power spectrum for each of DESI's redshift bins is presented in Figure \ref{fig:all-tracers-combined}. In both plots, the inferred value is shown at $k_{m,i}$, while the horizontal error bar represents the range of scales that contribute to this measurement.

To assess the quality of the fits, we compute the chi-squared ($\chi^2$) values for each tracer, as summarized in Table~\ref{tab:goodness-of-fit-table}. For the uncombined monopole and quadrupole moments, each $\chi^2$ value is based on 72 data points per tracer and 8 free parameters, shown in Table~\ref{tab:bestfit-params}. The $\Lambda$CDM model yields $\chi^2$ values ranging from 46 to 64 without inclusion of errorbars on $k_{m,i}$, indicating an overall goodness-of-fit comparable to that reported in the official Full-Shape galaxy clustering analysis \cite{desicollaboration2025desi2024vfullshape}. The poorest fits correspond to the largest $\chi^2$ values, found for the LRG1, LRG2, and QSO tracers, with deviations particularly notable in the quadrupole moment. Combining the galaxy monopole and quadrupole moments into a single set of 36 data points per tracer reduces the $\chi^2$ values to a range from 23 to 39, improving consistency with the $\Lambda$CDM model (Table~\ref{tab:goodness-of-fit-table}).

\begin{table}[ht]
\centering
\begin{tabular}{lcccccc}
\toprule
\textbf{Parameter} & \textbf{BGS} & \textbf{LRG1} & \textbf{LRG2} & \textbf{LRG3} & \textbf{ELG2} & \textbf{QSO} \\
\midrule
$b_{1}$                & 1.04   & 1.195 & 1.139 & 1.122 & 0.584 & 0.913 \\
$b_{2}$                & -1.4   & -1.39 & -0.80 & -0.7  & 0.14  & -0.64 \\
$b_{s}$                & 2.53   & 0.7   & 0.53  & -0.2  & 0.53  & 0.41 \\
$\alpha_{0}$           & -28    & 10    & -8.6  & 9.4   & 13    & 3.1 \\
$\alpha_{2}$           & 710    & 180   & 28    & 130   & 150   & 67 \\
$\alpha_{4}$           & -730   & -170  & 2.0   & -160  & -180  & -63 \\
$\textnormal{SN}_{0}$  & -0.78  & -0.71 & 0.57  & -0.28 & -0.15 & -0.0060 \\
$\textnormal{SN}_{2}$  & 0.084  & -3.4  & -1.7  & 0.42  & -0.35 & -0.42 \\
\bottomrule
\end{tabular}
\caption{Best-fit values of nuisance parameters, obtained via $\chi^2$ minimization using \texttt{velocileptors} (see Section~\ref{sec:Methodology-GC}). We fix $\mathrm{SN}_4 = 0$, since only monopole and quadrupole moments are included.}
\label{tab:bestfit-params}
\end{table}

\begin{table}[ht]
\centering

\begin{tabular}{llcccc}
\toprule
\textbf{} & \textbf{Tracer} & $\chi^2_{\Lambda \textnormal{CDM}}$(dof) & $\chi^2_{w_0w_a \textnormal{CDM}}$(dof) & \shortstack{p-value \\ ($\Lambda$CDM)} & \shortstack{p-value \\ ($w_0w_a$CDM)} \\
\midrule
\multirow{18}{*}{\shortstack[c]{Uncombined \\ Multipole \\ Fit}}
& \multirow{2}{*}{BGS}   & 46(64) & 46(64) & 0.96 & 0.96 \\
&                        & \textbf{38(64)} & \textbf{38(64)} & \textbf{$\sim$1} & \textbf{$\sim$1} \\
& \multirow{2}{*}{LRG1}  & 60(64) & 62(64) & 0.62 & 0.54 \\
&                        & \textbf{35(64)} & \textbf{38(64)} & \textbf{$\sim$1} & \textbf{$\sim$1} \\
& \multirow{2}{*}{LRG2}  & 59(64) & 59(64) & 0.67 & 0.66 \\
&                        & \textbf{35(64)} & \textbf{35(64)} & \textbf{$\sim$1} & \textbf{$\sim$1} \\
& \multirow{2}{*}{LRG3}  & 45(64) & 49(64) & 0.97 & 0.93 \\
&                        & \textbf{32(64)} & \textbf{34(64)} & \textbf{$\sim$1} & \textbf{$\sim$1} \\
& \multirow{2}{*}{ELG2}  & 57(64) & 52(64) & 0.74 & 0.87 \\
&                        & \textbf{44(64)} & \textbf{40(64)} & \textbf{0.97} & \textbf{$\sim$1} \\
& \multirow{2}{*}{QSO}   & 64(64) & 62(64) & 0.49 & 0.53 \\
&                        & \textbf{56(64)} & \textbf{55(64)} & \textbf{0.74} & \textbf{0.77} \\
& \multirow{2}{*}{All}   & 329(384) & 329(384) & 0.98 & 0.98 \\
&                        & \textbf{240(384)} & \textbf{239(384)} & \textbf{$\sim$1} & \textbf{$\sim$1} \\
& \multirow{2}{*}{All + CMB}  & 472(423) & 477(423) & 0.05 & 0.04 \\
&                             & \textbf{383(423)} & \textbf{388(423)} & \textbf{0.92} & \textbf{0.90} \\
& \multirow{2}{*}{All + CMB$_k$} & 332(423) & 333(423) & $\sim$1 & $\sim 1$ \\
&                                & \textbf{243(423)} & \textbf{243(423)} & \textbf{$\sim$1} & \textbf{$\sim$1} \\
\midrule
\multirow{18}{*}{\makecell[c]{Combined \\ Multipole \\ Fit}}
& \multirow{2}{*}{BGS}   & 28(28) & 28(28) & 0.46 & 0.46 \\
&                        & \textbf{22(28)} & \textbf{22(28)} & \textbf{0.77} & \textbf{0.77} \\
& \multirow{2}{*}{LRG1}  & 33(28) & 34(28) & 0.23 & 0.19 \\
&                        & \textbf{16(28)} & \textbf{19(28)} & \textbf{0.96} & \textbf{0.91} \\
& \multirow{2}{*}{LRG2}  & 34(28) & 34(28) & 0.20 & 0.19 \\
&                        & \textbf{19(28)} & \textbf{19(28)} & \textbf{0.91} & \textbf{0.90} \\
& \multirow{2}{*}{LRG3}  & 23(28) & 26(28) & 0.71 & 0.57 \\
&                        & \textbf{13(28)} & \textbf{16(28)} & \textbf{$\sim$1} & \textbf{0.96} \\
& \multirow{2}{*}{ELG2}  & 37(28) & 32(28) & 0.12 & 0.29 \\
&                        & \textbf{23(28)} & \textbf{16(28)} & \textbf{0.74} & \textbf{0.96} \\
& \multirow{2}{*}{QSO}   & 39(28) & 39(28) & 0.08 & 0.09 \\
&                        & \textbf{32(28)} & \textbf{32(28)} & \textbf{0.26} & \textbf{0.27} \\
& \multirow{2}{*}{All}   & 58(28) & 63(28) & $<$ 0.01 & $<$ 0.01 \\
&                        & \textbf{34(28)} & \textbf{35(28)} & \textbf{0.19} & \textbf{0.17} \\
& \multirow{2}{*}{All + CMB}  & 201(67) & 212(67) & $<$0.01 & $<$0.01 \\
&                             & \textbf{177(67)} & \textbf{183(67)} & \textbf{$<$0.01} & \textbf{$<$0.01} \\
& \multirow{2}{*}{All + CMB$_k$} & 61(67) & 67(67) & 0.67 & 0.48 \\
&                                & \textbf{38(67)} & \textbf{39(67)} & \textbf{$\sim$1} & \textbf{$\sim$1} \\
\bottomrule
\end{tabular}

\caption{Goodness of fit test for each tracer and for all tracers in combination with CMB, producing a global fit. The table shows $\chi^2$ values, degrees of freedom and the corresponding p-values for both $\Lambda$CDM and $w_0w_a$CDM cosmological models. Bold entries indicate fits including $k$-errorbars from the galaxy power spectrum in the galaxy covariance, while CMB$_k$ indicates inclusion of $k$-errorbars from the CMB power spectrum in the CMB covariance.}
\label{tab:goodness-of-fit-table}
\end{table}

For the combined data across all galaxy tracers, assuming the datasets are uncorrelated, we obtain a total $\chi^2 = 329$ with $384$ dof, reinforcing that the $\Lambda \textnormal{CDM}$ model provides a statistically acceptable fit. 
Focusing on the CMB data alone, the initial fit yields $\chi^2 = 148$ for $39$ datapoints when errors in $k$ and correlations between $k$-modes are neglected. This is the approach used in \cite{Chabanier_2019}. However, as can be seen from Figure \ref{fig:final-plot-desi-all}, the $k$-errors are substantial for CMB datapoints and remain non-negligible for the combined galaxy clustering measurements at the smallest scales and so should not be neglected. Incorporating horizontal errors into the covariance by estimating the squared slope of the theoretical matter power spectrum, multiplying it by the variance in $k$, and adding this to the diagonal elements, we then rescale the full covariance matrix accordingly (without introducing additional $k$-correlations). This procedure reduces the $\chi^2$ for CMB-alone significantly to $3.5$, suggesting that the uncertainties are overestimated but highlighting the importance of including $k$-errors. The low $\chi^2$ value could indicate that our use of 20\%–80\% intervals to define $k$-uncertainties is overly conservative, or that correlations between $k$-bins may also need to be accounted for. A more formal treatment would involve constructing a likelihood for each data point, marginalised over the unknown true $k$-value. However, this approach does not yield a standard $\chi^2$ expression and as this work is focused on galaxy data, we leave further investigation into the CMB aspects for future studies. Furthermore, we propagate the $k_{m,i}$ uncertainties from galaxy clustering measurements in the same way as for the CMB. This substantially improves the individual tracer fits and reduces the global $\chi^2$ to 240 for the uncombined multipoles and 34 for the combined monopole–quadrupole case. The reduction in $\chi^2$ arises because the inverse-variance combination effectively marginalizes over discrepancies between the monopole and quadrupole when they are treated as independent observables. Owing to the strong correlation between the two multipoles, this compression removes modes that are particularly sensitive to residual modeling uncertainties, leading to an apparently better goodness of fit due to the reduced information content.

In the global fit---combining CMB and galaxy clustering data---we obtain a total $\chi^2 = 472$ for 423 dof when treating the monopole and quadrupole moments of each tracer separately. If, instead, the monopole and quadrupole are combined using a definition inverse-variance-weighted average for the correlated data, the total reduces to $\chi^2 = 201$ for 67 dof. In both cases, the CMB contributes 39, while galaxy clustering contributes either $6 \times (72-8)$ (uncombined) or $36-8$ (averaged) datapoints. These fits assume no correlation between low and high-$\ell$ multipoles in the CMB TT spectrum, no $k$-mode correlations, and neglect uncertainties in $k$-binning. Accounting for $k$-uncertainty in the CMB, the total $\chi^2$ increases to 332 for the global fit with separate galaxy clustering monopole and quadrupole moments, and to 61 for the global fit using combined galaxy moments. Finally, accounting for the horizontal errorbars of both CMB and galaxy clustering datapoints, results a $\chi^2 = 38$ for 67 dof as the goodness-of-fit for the result presented in Figure~\ref{fig:final-plot-desi-all}. The number of data points remains the same as in the corresponding fits without $k$-uncertainty. Notably, the TT portion of the CMB angular power spectrum contributes the largest discrepancy, providing the weakest fit to $\Lambda$CDM. These results indicate that the $\Lambda$CDM model provides a reasonable description of the global data within the confines of the $\Lambda$CDM cosmology.

The findings of the DESI Collaboration \cite{desicollaboration2024novVII, desicollaboration2025ii, desicollaboration2024desi1} demonstrate that DESI Y1 data plays a crucial role in constraining the $w_0w_a\text{CDM}$ cosmology. When DESI Full-Shape and BAO measurements are combined with external datasets --- namely CMB and SN Ia data --- the preference for a time-evolving dark energy equation of state is strengthened, pointing to deviations from the standard $\Lambda\text{CDM}$ model.

\begin{figure}
\centering
    \includegraphics[width=\textwidth]{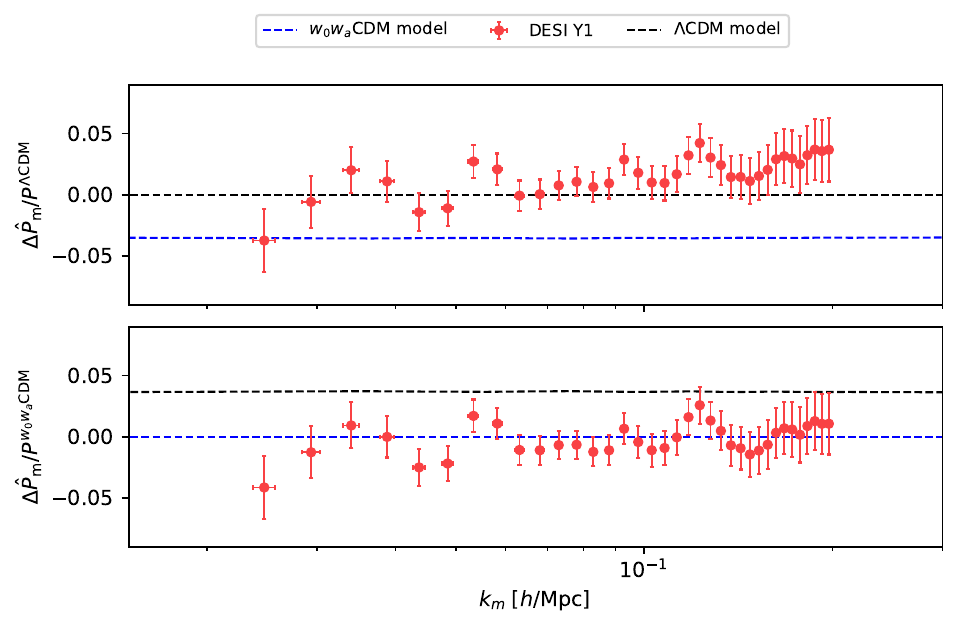}
    \caption{Residuals of the present-day matter power spectrum recovered from DESI galaxy clustering data. The top panel shows the residuals of the matter power spectrum inferred from the data, assuming a fixed $\Lambda$CDM cosmology, relative to the corresponding $\Lambda$CDM theoretical prediction of matter power spectrum. The bottom panel shows the residuals of results under assumption of a fixed $w_0w_a$CDM cosmology, again relative to its own theoretical prediction. Dashed lines indicate the shift between cosmological models: the blue dashed line represents the residuals for theoretical prediction of matter power spectrum at $z=0$ in $w_0w_a$CDM background, while the black dashed line represents the prediction under $\Lambda$CDM. Note that the different $k$-bins of the recovered power spectrum are highly correlated, so apparent deviations from the model in the plot may not reflect true statistical significance.}
    \label{fig:difference in cosmologies}
\end{figure}

In this context, comparing the reconstructed matter power spectra under fixed $\Lambda$CDM and $w_0w_a$CDM backgrounds provides a valuable consistency test for the sensitivity of the DESI Y1 Full-Shape results to the assumed cosmology. Fixing the background to the $w_0w_a$CDM model produces a matter power spectrum that closely resembles the fiducial $\Lambda\text{CDM}$ case. This similarity arises from minor shifts in the best-fit galaxy power spectrum nuisance parameters when switching between the two models, suggesting that these parameters absorb much of the cosmological differences. However, some residual BAO features remain visible in both cases, indicating that the match between data and theory is not exact. These effects are illustrated in Figure~\ref{fig:difference in cosmologies}, which shows residuals in the inferred matter power spectra at redshift $z=0$ --- recovered from the combined monopole and quadrupole moments --- under both cosmological backgrounds. Notably, an upward trend in the residuals appears in the $\Lambda$CDM case but not in the $w_0w_a$CDM case, suggesting that some sensitivity to the assumed cosmological model persists. This distinction is not necessarily reflected in the $\chi^2$ values in Table~\ref{tab:goodness-of-fit-table}. Here, in the combined multipole case when horizontal errors are neglected, the tracers contribute to an increased $\chi^2$ for both cosmologies, with the goodness-of-fit values favouring $\Lambda$CDM cosmology. We identify that this increase reflects not only the quality of the fit but also correlations between measurements, which may share systematics incorporated into the covariance matrix. Note that, unlike in the individual galaxy power spectra, the $k$ bins of the reconstructed matter power spectrum are highly correlated. The rise in $\chi^2$ is primarily driven by correlations from LRG1, LRG2, and ELG2, with additional contributions from BGS in the $\Lambda$CDM case. However, inclusion of $k_{m,i}$ errorbars in the calculation of goodness-of-fit improves the fits for both cosmologies, without any significant preference for one over the other. Repeating this analysis with future DESI data would be particularly valuable, especially given the apparent differences between the cosmological models in the residual trends shown in Figure~\ref{fig:difference in cosmologies}. While these differences are evident visually, they are not strongly reflected in the corresponding $\chi^2$ and $p$-values reported for the $w_0w_a$CDM background in Table~\ref{tab:goodness-of-fit-table}. This suggests that either a more precise dataset or a more sensitive modelling statistic may be required to fully leverage this analysis.

When treating the monopole and quadrupole moments of galaxy clustering separately, the ELG2 and the QSO tracers show a reduction in $\chi^2$ under the $w_0w_a$CDM model. In contrast, LRG1 and LRG3 shows an increase in $\chi^2$. After combining the multipole moments of each tracer, the reduction in $\chi^2$ for ELG2 remains pronounced when transitioning from $\Lambda$CDM to $w_0w_a$CDM. All other tracers do not show significant changes in $\chi^2$ statistic when switching from one cosmology to the other after inclusion of errors in $k$. We find a total $\chi^2 = 329$ with 384 dof for the uncombined multipole case and $\chi^2 = 63$ with 28 dof for the combined galaxy multipole measurements in $w_0w_a$CDM background, further supporting the conclusion that DESI Y1 Full-Shape data alone is consistent with both the $\Lambda$CDM and $w_0w_a$CDM models, showing no significant preference for either cosmology in this work. In the global fit for $w_0w_a$CDM, neglecting CMB correlations and uncertainties in $k$, we obtain $\chi^2 = 477$ for 423 dof when treating galaxy clustering multipoles separately and $\chi^2 = 212$ for a total $67$ dof when compressing galaxy clustering multipoles into a single measurement. Upon including the errors in $k$ for both CMB and galaxy clustering, we get $\chi^2 = 333$ and $\chi^2 = 39$ for the same number of datapoints in uncombined and combined galaxy multipole case respectively. 
Overall, these results indicate that DESI data alone shows good agreement with both the $\Lambda$CDM and $w_0w_a$CDM models, highlighting the importance of incorporating additional datasets to more effectively identify any potential discrepancies in the analysis of matter power spectrum inference.

\section{Conclusion}

In this work, we reconstructed the 3D matter power spectrum at redshift $z=0$ using Cosmic Microwave Background observations and DESI Year 1 galaxy clustering data. By combining early-Universe constraints from the CMB with late-time galaxy clustering measurements, we updated the inferred matter power spectrum across scales ranging from \( k = 2 \times 10^{-4} \, h\,\mathrm{Mpc}^{-1} \) to \( k = 0.2 \, h\,\mathrm{Mpc}^{-1} \) and explored the consistency of observations from DESI within $\Lambda \textnormal{CDM}$ and $w_0w_a \textnormal{CDM}$. Building on previous work, we applied the same reconstruction of matter power spectrum approach to CMB temperature anisotropies and polarization measurements as in \cite{Chabanier_2019}, while implementing an alternative data binning strategy in the analysis of the DESI DR1 Full-Shape power spectrum \cite{desicollaboration2024novVII} across the $k$ range. Our methodology incorporated LPT within the \texttt{velocileptors} framework to model galaxy bias and account for non-linear effects on intermediate scales, enabling accurate modelling of the galaxy power spectrum (see \cite{maus2024comparisoneffectivefieldtheory} for a detailed model comparison).

Our results indicate that the DESI Year 1 Full-Shape data—both for all galaxy tracers combined and individual tracers—remain consistent with the $\Lambda \text{CDM}$ model. The computed statistics further confirm a satisfactory fit across all tracers. Motivated by emerging hints of evolving Dark Energy, we also tested the compatibility of galaxy clustering data with the $w_0w_a\text{CDM}$ model, assessing both visual trends and statistical goodness of fit. Our findings show that both $\Lambda \text{CDM}$ and $w_0w_a\text{CDM}$, provide reasonable fits to the DESI Year 1 Full-Shape data, with no strong preference for one cosmology over the other based on chi-squared values. Beyond conventional Bayesian parameter inference, our analysis introduces an alternative approach to testing the consistency of galaxy clustering data with cosmological models. By directly tracing the underlying matter power spectrum across different observations, scales, and redshifts, this method offers a transparent and effective compatibility test of the cosmological model. 

However, this approach comes with certain limitations. It is not model-independent and relies on assumptions such as the adequacy of Kaiser modelling in extracting $k_{m,i}$ for galaxy clustering, a fixed background cosmology, and the validity of perturbative modelling. As a result, it is more suitable as a qualitative consistency check rather than a rigorous statistical test. Furthermore, the method draws on inference techniques developed for different probes, each with its own systematics and modelling assumptions, which must be carefully accounted for when combining results.

Despite these caveats, the approach complements existing techniques for testing cosmological models and provides a flexible framework for exploring alternative scenarios. For future work, we aim to extend our analysis by incorporating Ly-$\alpha$ forest measurements from DESI DR1. With the advent of high-precision cosmology, approaches such as ours will be essential in cross-checking results and deepening our understanding of the underlying model of the Universe.

\section*{Data availability}
Data from the plots in this paper will be made available on Zenodo\footnote{} as part of DESI’s Data Management Plan. The data used in this analysis is public along with Data Release 1 of DESI of 2025\footnote{Details can be found here: \hyperlink{https://data.desi.lbl.gov/doc/releases/}{https://data.desi.lbl.gov/doc/releases/}}.

\acknowledgments
We would like to thank Nathalie Palanque-Delabrouille and Krishna Naidoo for serving as internal reviewers of this work and providing useful feedback. We also thank Arnaud De Mattia for help with the \texttt{desilike} framework. RC would also like to thank Ot\'avio Alves and Xavier Pritchard for useful discussions.
RC is supported by a UK Science and Technology Facilities Council (STFC) studentship and the University of Sussex. EM acknowledges support from an STFC Ernest Rutherford Fellowship, with grant reference ST/W004755/1.

This material is based upon work supported by the U.S. Department of Energy (DOE), Office of Science, Office of High-Energy Physics, under Contract No. DE–AC02–05CH11231, and by the National Energy Research Scientific Computing Center, a DOE Office of Science User Facility under the same contract. Additional support for DESI was provided by the U.S. National Science Foundation (NSF), Division of Astronomical Sciences under Contract No. AST-0950945 to the NSF’s National Optical-Infrared Astronomy Research Laboratory; the Science and Technology Facilities Council of the United Kingdom; the Gordon and Betty Moore Foundation; the Heising-Simons Foundation; the French Alternative Energies and Atomic Energy Commission (CEA); the National Council of Humanities, Science and Technology of Mexico (CONAHCYT); the Ministry of Science, Innovation and Universities of Spain (MICIU/AEI/10.13039/501100011033), and by the DESI Member Institutions: \url{https://www.desi.lbl.gov/collaborating-institutions}. Any opinions, findings, and conclusions or recommendations expressed in this material are those of the author(s) and do not necessarily reflect the views of the U. S. National Science Foundation, the U. S. Department of Energy, or any of the listed funding agencies.

The authors are honored to be permitted to conduct scientific research on I'oligam Du'ag (Kitt Peak), a mountain with particular significance to the Tohono O’odham Nation.

\bibliographystyle{JHEP}
\bibliography{literature.bib}

\appendix

\section{Validation of $k$ range choices using mocks}
\label{appendix}

We explicitly validated our chosen $k$-range ($0.02\,h/{\text{Mpc}} < k < 0.2\,h/{\text{Mpc}}$) using 25 AbacusSummit mock realizations of the BGS tracer at low redshift ($0.1 < z < 0.4$). This tracer is the most sensitive to nonlinear effects among the samples we consider, as structure growth is most advanced at low redshift. Therefore, if nonlinearities were leaking into the reconstructed linear matter power spectrum, they would be most apparent in the BGS sample. 

To test the robustness of our method when the cosmology in question differs from the true cosmology of the mocks, we applied our analysis to the AbacusSummit mocks (produced under DESI's fiducial $\Lambda$CDM) while adopting a shifted $\Lambda$CDM with parameters $n_s = 0.969$, $\Omega_m = 0.3117$, $h = 0.676$, $\ln(10^{10} A_s) = 3.035$, $\tau = 0.048$, $\omega_b = 0.0225$, and $\omega_{\rm cdm} = 0.1193$, with all remaining parameters fixed to the DESI's fiducial values. These values are within the $1\sigma$ uncertainties of the Planck $\Lambda$CDM constraints.

\begin{figure}
    \centering
    \includegraphics[width=0.8\linewidth]{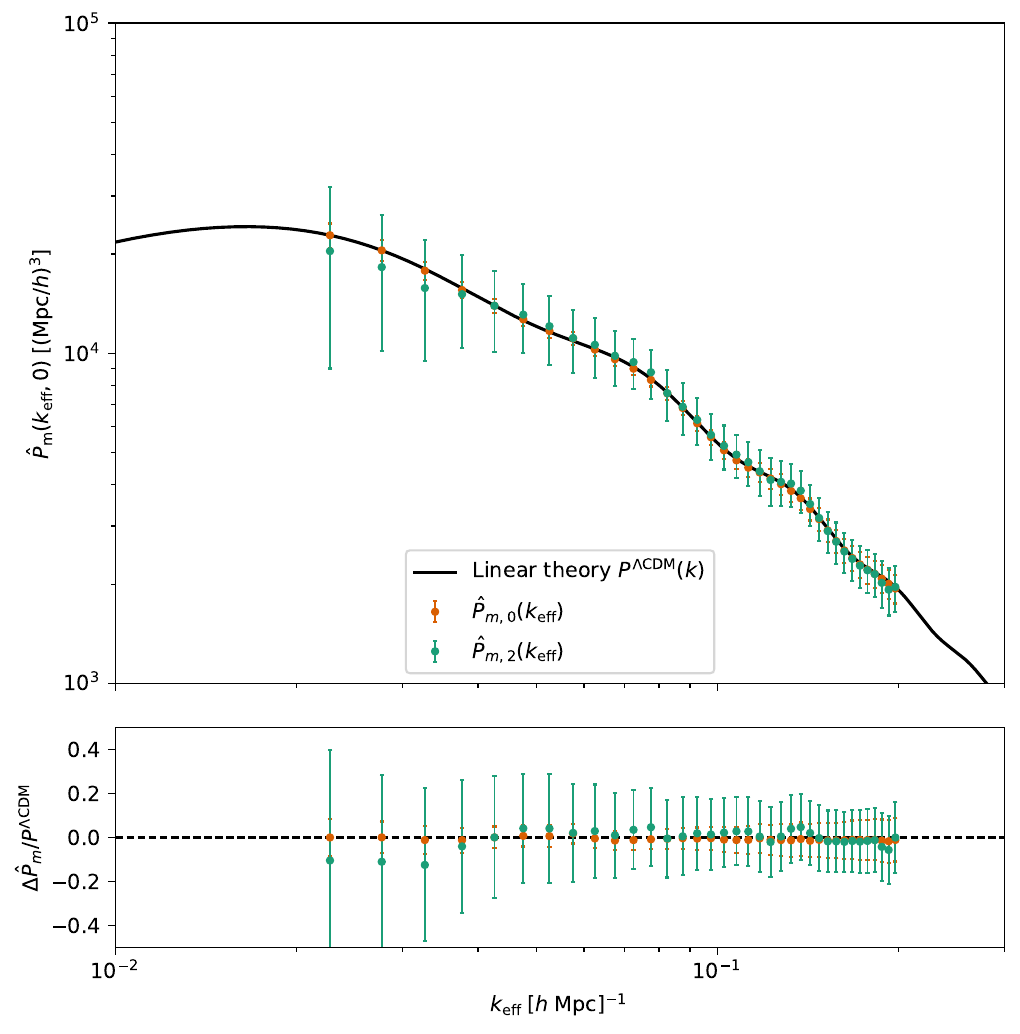}
    \caption{Linear matter power spectrum inferred from 25 Abacus Summit mock realisations of the BGS tracer ($0.1<z<0.4$). The black line shows the theoretical linear matter power spectrum at $z=0$, while the points show the inference result. The bottom panel shows the residual of recovered datapoints. Since any apparent nonlinearities would be most prominent at low redshift, this represents the most stringent case among our tracers. The agreement demonstrates that our method robustly recovers the linear spectrum over the $k$-range used in the main analysis.}
    \label{fig:mock-k-range-test}
\end{figure}

In Figure~\ref{fig:mock-k-range-test}, we show the mean inferred matter power spectrum from the monopole and quadrupole measurements of 25 mocks, compared to the theoretical linear matter spectrum at redshift $z=0$. Over the full $k$ range used in our main analysis, the reconstruction accurately recovers the linear theory prediction within the uncertainties, both for the mock and for the data. This demonstrates that any residual nonlinear corrections are negligible for our chosen scales, and confirms that our methodology robustly recovers the linear power spectrum even when the fiducial cosmology differs slightly from the true cosmology of the mocks. We therefore conclude that our methodology is robust in the $k$-range adopted throughout the paper. 

\end{document}